\documentclass[twocolumn, secnumarabic, amsmath, amssymb, nobibnotes, aps, pra]{revtex4-2}

\usepackage{graphicx}
\usepackage{dcolumn}
\usepackage{bm}

\usepackage{upgreek}
\usepackage{amsmath}
\usepackage{braket}
\usepackage{xcolor}
\usepackage{ulem}
\usepackage{soul}
\usepackage[colorlinks=True, linkcolor=blue, urlcolor=blue]{hyperref}

\begin{document}

\title{Robust Generation of Topological Biphoton Mode via Adiabatic Passage}

\author{Jaesung Lim}
\author{Jihwan Kim}
\author{Dong-Gil Im}
\author{Kyungdeuk Park}
\author{Dongkyu Kim}
\author{Yonggi Jo}
\author{Yong Sup Ihn}
\email{yong0862@add.re.kr}
\affiliation{Agency for Defense Development, Daejeon 34186, Korea}
\date{\today}

\begin{abstract}
  Topological waveguide arrays support robust mode propagation in the presence of fabrication imperfections, providing a significant advantage for on-chip quantum information processing.
  However, this robustness does not fully extend to nonlinear biphoton generation.
  Structural disorder can enhance the excitation of non-topological biphoton modes during nonlinear interactions, which degrades the quantum properties of the generated state.
  To overcome this limitation, we propose an adiabatic passage that connects an isolated site to a topological defect array.
  By initiating the nonlinear process in a strongly isolated regime, nonlinear coupling to unwanted modes is effectively suppressed, thereby preserving the Schmidt number of the generated state.
  The subsequent adiabatic connection facilitates the high fidelity transfer of the generated biphoton into the topological biphoton mode.
  Our numerical simulations demonstrate that, unlike conventional topological structures, the adiabatic scheme maintains both high biphoton fidelity and a unit Schmidt number in the presence of waveguide gap disorder.
  Furthermore, we show that this robustness extends to path entangled NOON states, achieving a near-unity quantum interference visibility.
  Our approach provides a practical design strategy for disorder-tolerant integrated quantum photonic devices.
\end{abstract}
\maketitle


\section{\label{sec1}Introduction}
Nonclassical light serves as a key resource for quantum information processing due to its inherently weak interaction with the environment, which leads to low decoherence~\cite{wang2020integrated, moody20222022, luo2023recent, silverstone2016silicon}.
Integrated quantum photonics, made possible by advanced fabrication technologies, further enhances the functionality of quantum communication~\cite{sibson2017chip, zhang2019integrated} and quantum computation~\cite{politi2009shor, psiquantum2025manufacturable} while improving overall system compactness and stability.
On-chip biphoton generation can be implemented using $\chi^{(2)}$ or $\chi^{(3)}$ nonlinear waveguides~\cite{dutt2024nonlinear}, where the biphoton state is described in a spatial basis labeled by the waveguide indices occupied by each photon.
The coupling between the nonlinear waveguides provides versatile spatial mode control capabilities, including tunable generation of entangled states~\cite{silverstone2014chip, jin2014chip, setzpfandt2016tunable}, and manipulation of spatial correlations through quantum walks~\cite{solntsev2012spontaneous, solntsev2012photon, solntsev2014generation}, all without requiring additional biphoton generation setups.
However, the spatial mode profiles of biphotons are highly sensitive to fabrication imperfections, such as deviations in waveguide gaps, which can ultimately degrade device performance.

Topological photonics offers a promising strategy for overcoming such limitations by employing topological modes that appear at the interface between lattices with different topological phases~\cite{lu2014topological, ozawa2019topological}.
A representative one-dimensional model that supports topological modes is the Su--Schrieffer--Heeger (SSH) lattice~\cite{su1979sitons}, a dimerized chain formed by alternating intracell and intercell hopping strengths.
When the ratio of these two hopping amplitudes exceeds unity, the topological invariant of the lattice---the Zak phase---undergoes a transition~\cite{zak1989berry}.
Bringing two SSH lattices with distinct topological phases together results in the emergence of the topological mode localized at the interface~\cite{asboth2016short}.
This SSH model can be implemented in waveguide arrays to enable robust field propagation~\cite{cheng2015topologically, blanco2016topological} and routing~\cite{song2020robust} against fabrication imperfections, as long as the chiral symmetry of the Hamiltonian is preserved and the bulk topological invariant remains unchanged.

Recent studies have shown that topological modes can also provide robustness for the spatial profiles of biphotons~\cite{yan2021quantum, smirnova2020nonlinear, hashemi2025topological}.
In SSH waveguide lattices, topological biphoton states, in which both the signal and idler are supported by the topological mode, can retain protected features---zero amplitude on every other waveguide and unchanged propagation constant---under gap disorder.
These properties offer an additional degree of freedom for disorder-resistant nonclassical interference~\cite{blanco2018topological, wang2019topologically}, high dimensional entanglement~\cite{doyle2022biphoton, zakeri2026high}, and hyperentanglement~\cite{bergamasco2021generation}.
Despite these advantages, a fundamental challenge remains: in nonlinear generation, the modal overlap between the topological biphoton mode and the pump field is neither perfect nor protected.
Fabrication-induced waveguide gap disorder slightly deforms the topological mode, thereby reducing the modal overlap and increasing the generation of unintended non-topological biphoton modes~\cite{bergamasco2019generation}.
This degradation occurs even when the bulk topological invariant remains unchanged, indicating that the robustness of the topological mode is not fully carried over to the nonlinear generation.

To address this challenge, we propose a robust scheme for generating topological biphoton modes using an adiabatic passage that enables high fidelity excitation of topological modes~\cite{longhi2019topological, liu2024perfect}.
Adiabatic passage has previously been used for excitation and pumping of topological modes in the linear regime.
Our strategy instead applies it directly to the nonlinear generation and stabilizes the modal overlap of the nonlinear coupling against gap disorder.
Specifically, we consider a waveguide array with an isolated central site, adiabatically linked to a central-defect SSH lattice (Fig.~\ref{fig1}(a)).
By initiating the nonlinear process within this isolated regime, unwanted nonlinear interactions are suppressed through strong mode confinement, preserving the biphoton Schmidt number (SN).
The generated state is then transferred into the topological biphoton mode with high fidelity via the adiabatic connection.
Numerical simulations demonstrate that this approach significantly enhances robustness against waveguide gap disorder compared to conventional topological structures.
By examining a control case in which the adiabatic passage is applied to a trivial defect array, we verify that the enhanced robustness arises from the complementary action of two factors: strong spatial isolation at the initial stage, which suppresses unwanted biphoton generation, and the topological mode, which preserves the spectral isolation needed for stable adiabatic transfer.
Furthermore, we extend this method to path entangled NOON states, achieving an exceptional quantum interference visibility of 99.9$\%$.
Our results provide a disorder-resilient and versatile scheme for integrated quantum information processing.


\section{\label{sec2}Results}
We consider a one-dimensional silicon waveguide array implemented on a silicon-on-insulator (SOI) platform.
The source and device parameters are adopted from the experimental demonstration \cite{blanco2018topological}.
A continuous-wave pump laser with a central wavelength of \(1550\,\mathrm{nm}\) is injected into the waveguide array.
Biphoton generation occurs via spontaneous four-wave mixing (SFWM) driven by the third-order optical nonlinearity of silicon, where two pump photons are annihilated to produce a signal--idler photon pair.
We assume the use of narrowband filtering, which restricts the collection bandwidth and allows the biphoton state to be described solely in the spatial basis.
The signal and idler wavelengths are set to \(1545\,\mathrm{nm}\) and \(1555\,\mathrm{nm}\), respectively.

Each waveguide has a width of \(w = 450\,\mathrm{nm}\) and a height of \(h = 220\,\mathrm{nm}\).
The refractive indices of the silicon core and silicon dioxide substrate are taken to be \(n_{\mathrm{Si}} = 3.48\) and \(n_{\mathrm{SiO}_{2}} = 1.47\), respectively.
The SSH lattice is realized by alternately arranging short and long air gaps with widths of \(g_\mathrm{s} = 173\,\mathrm{nm}\) and \(g_\mathrm{l} = 307\,\mathrm{nm}\), respectively.
The array consists of 121 waveguides, indexed 0 through 120.
A ``long-long'' defect is formed by placing long gaps on both sides of the central waveguide, producing a single central topological defect mode.
The Hamiltonian of the topological lattice is constructed using nearest-neighbor coupling constants extracted from finite-element method (FEM) simulations and is given by
\begin{equation}
  H_{\mathrm{SSH}}
  = \sum_{i=0}^{119}
  c_i\left(\ket{i}\bra{i+1}+ \ket{i+1}\bra{i} \right).
  \label{eq1}
\end{equation}
Here, \(\ket{i}\) denotes the mode localized at the \(i\)-th waveguide, and \(c_i\) represents the coupling constant between the modes of the \(i\)-th and \((i+1)\)-th waveguides.
For \(i \le 59\), \(c_i\) corresponds to the short (long) gap coupling when \(i\) is even (odd); this assignment is reversed for \(i \geq 60\).

\begin{figure}[t]
  \centering
  \includegraphics[width=0.5\textwidth]{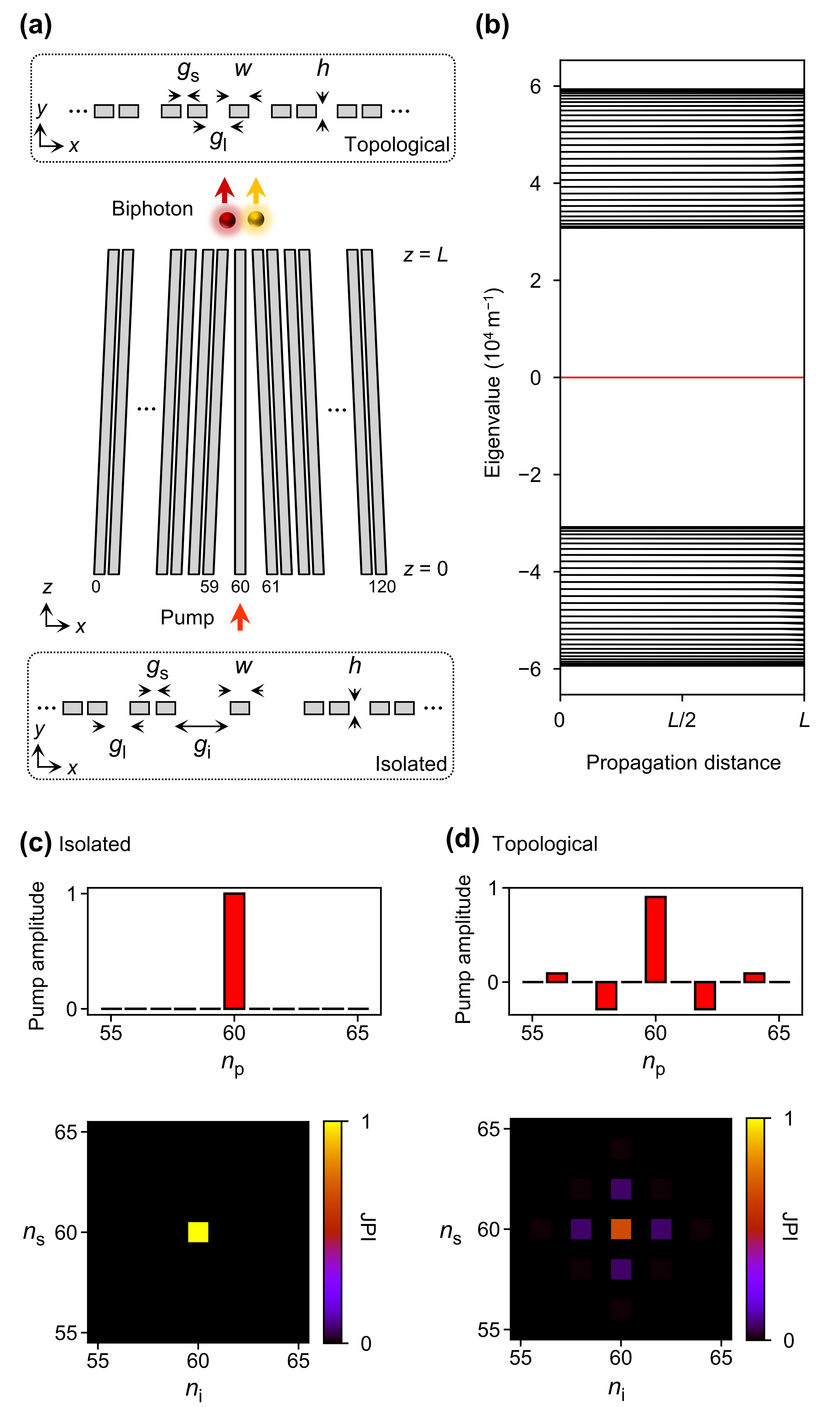}
  \caption{\label{fig1} Concept of topological biphoton generation via adiabatic passage in a silicon waveguide array.
    (a) Schematic of the adiabatic connection from an isolated-waveguide array to a topological lattice.
    (b) Variation of the eigenvalue spectrum during propagation, indicating that the separation between the bulk bands and the midgap state is preserved.
  (c, d) Pump amplitude and biphoton JPI, both supported by the midgap mode of (c) the isolated-waveguide lattice and (d) the topological lattice.}
\end{figure}

The waveguide array containing the isolated central site (isolated-waveguide array) is formed by introducing isolation gaps \(g_\mathrm{i}\) around the central waveguide, instead of the long gaps used in the topological lattice.
To ensure isolation, $g_\mathrm{i}$ is set to \(930\,\mathrm{nm}\), corresponding to the next-nearest-neighbor distance, which results in a coupling constant that is only $0.6\%$ of the long gap coupling.
The connection between the isolated-waveguide array and the topological lattice is implemented by linearly varying the waveguide gaps over the total device length $\mathit{L}$, as illustrated in Fig.~\ref{fig1}(a).
The midgap eigenvalue remains well separated from the bulk bands throughout the device, as shown in Fig.~\ref{fig1}(b).
This behavior is particularly favorable for adiabatic evolution and suppresses mode mixing.

\begin{figure*}[t]
  \centering
  \includegraphics[width=1\textwidth]{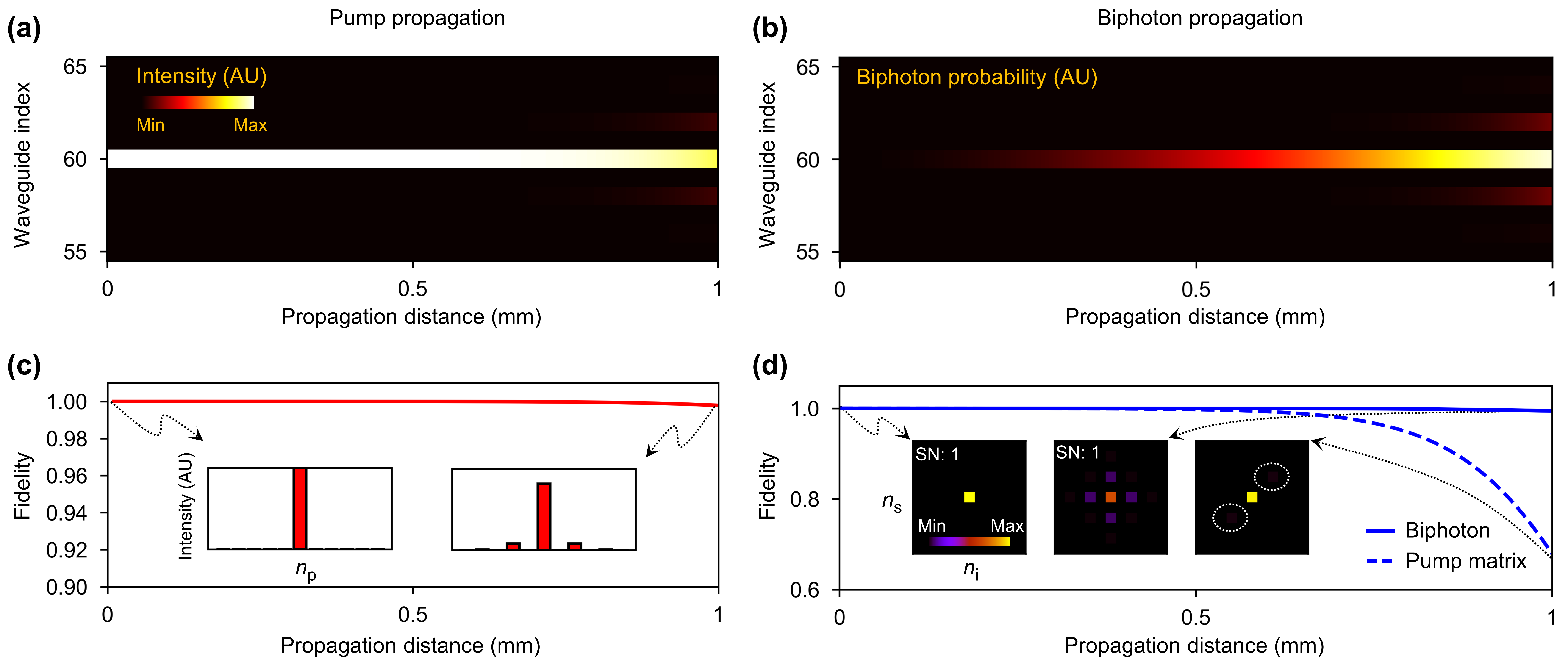}
  \caption{\label{fig2} Adiabatic evolution of the pump and biphoton states for a device length of \(1.0\,\mathrm{mm}\).
    (a) Evolution of the pump intensity along the propagation distance.
    (b) Biphoton probability distribution during the adiabatic passage.
    (c) Fidelity of the propagating pump field with respect to the midgap mode, confirming high fidelity adiabatic transfer. The insets show the pump intensity profiles at the input and output facets.
  (d) Fidelity of the evolving biphoton state (solid line) and the normalized pump matrix defined in Eq.~\eqref{eq6} (dashed line), both relative to the midgap biphoton mode. The insets display the JPI profiles and the corresponding SN.}
\end{figure*}

The pump is injected into the isolated site and generates biphotons while it evolves through the adiabatic passage.
The propagation of the pump is governed by the coupled mode equation
\begin{equation}
  i\,\frac{d}{dz}\ket{A} = H\ket{A},
  \label{eq2}
\end{equation}
where \(H\) is the lattice Hamiltonian and $\ket{A} = (A_\mathrm{0}, \ldots, A_{N-1})^{T}$ denotes the field amplitudes of the $N$ waveguides.
The generated biphoton state can be expressed in the spatial basis as $\ket{\psi} = \sum_{j,k} \psi_{jk}\,\ket{1_j 1_k}_{\mathrm{s,i}}$, where $\ket{1_j 1_k}_{\mathrm{s,i}}$ represents the signal and idler photons located in waveguides \(j\) and \(k\), respectively, and $\psi_{jk}$ is the joint path amplitude (JPA).
Fig.~\ref{fig1}(c) presents the spatial distributions of the pump amplitude and the joint path intensity (JPI) of the biphoton, where the pump, signal, and idler are all supported by the midgap mode in the isolated-waveguide lattice.
\(n_{\mathrm{p,s,i}}\) denotes the waveguide indices of the pump, signal, and idler, respectively.
Fig.~\ref{fig1}(d) shows the corresponding distributions in the topological lattice.
In the isolated-waveguide array, the midgap mode is strictly confined to the central waveguide, and the biphoton inherits this localized profile.
On the other hand, the midgap mode in the topological array exhibits a distribution confined to the even-indexed waveguides, and the biphoton mode also retains this feature.

A biphoton generated in the midgap signal and idler modes (the midgap biphoton mode) subsequently evolves into the topological biphoton mode through the adiabatic connection.
The evolution of the biphoton is governed by
\begin{equation}
  i\,\frac{\partial \psi_{jk}}{\partial z}
  = \sum_{l} \left[ H^{\mathrm{s}}_{jl}\psi_{lk} + H^{\mathrm{i}}_{kl}\psi_{jl} \right] + \gamma A_j^{2}(z)\,\delta_{jk},
  \label{eq3}
\end{equation}
where $H^{\mathrm{s}}_{jl}$ and $H^{\mathrm{i}}_{kl}$ are the matrix elements of the lattice Hamiltonians for the signal and idler, $A_j^{2}(z)$ represents the squared pump field amplitude in the $j$-th waveguide, \(\delta_{jk}\) is the Kronecker delta, and $\gamma$ is the nonlinear parameter of the silicon waveguide.
The first and second terms on the right-hand side describe the evolution of the signal and idler photons, respectively.
Differences in the propagation constants of the pump, signal, and idler---arising from their distinct wavelengths---are incorporated through diagonal elements in the lattice Hamiltonian.
The final term represents biphoton generation via SFWM.
$\gamma$ is given by \(\gamma = 2\pi n_{\mathrm{2}} /{\lambda_{\mathrm{pump}} A_{\mathrm{eff}}}\), where \( n_{2}\) is the nonlinear refractive index, \(\lambda_{\mathrm{pump}}\) is the pump wavelength, and \(A_{\mathrm{eff}}\) is the effective mode area.
Combined with the reported nonlinear refractive index of the silicon, \(n_{\mathrm{2}} = 4.5 \times 10^{-18}\,\mathrm{m^{2}\,W^{-1}}\)~\cite{dinu2003third}, and the calculated effective mode area, this yields \(\gamma = 285.47\,\mathrm{m^{-1}\,W^{-1}}\), which is used in all numerical simulations.
Self-phase modulation is neglected, as its contribution is negligible for the milliwatt-level pump powers considered in this study.

\begin{figure*}[t]
  \centering
  \includegraphics[width=1\textwidth]{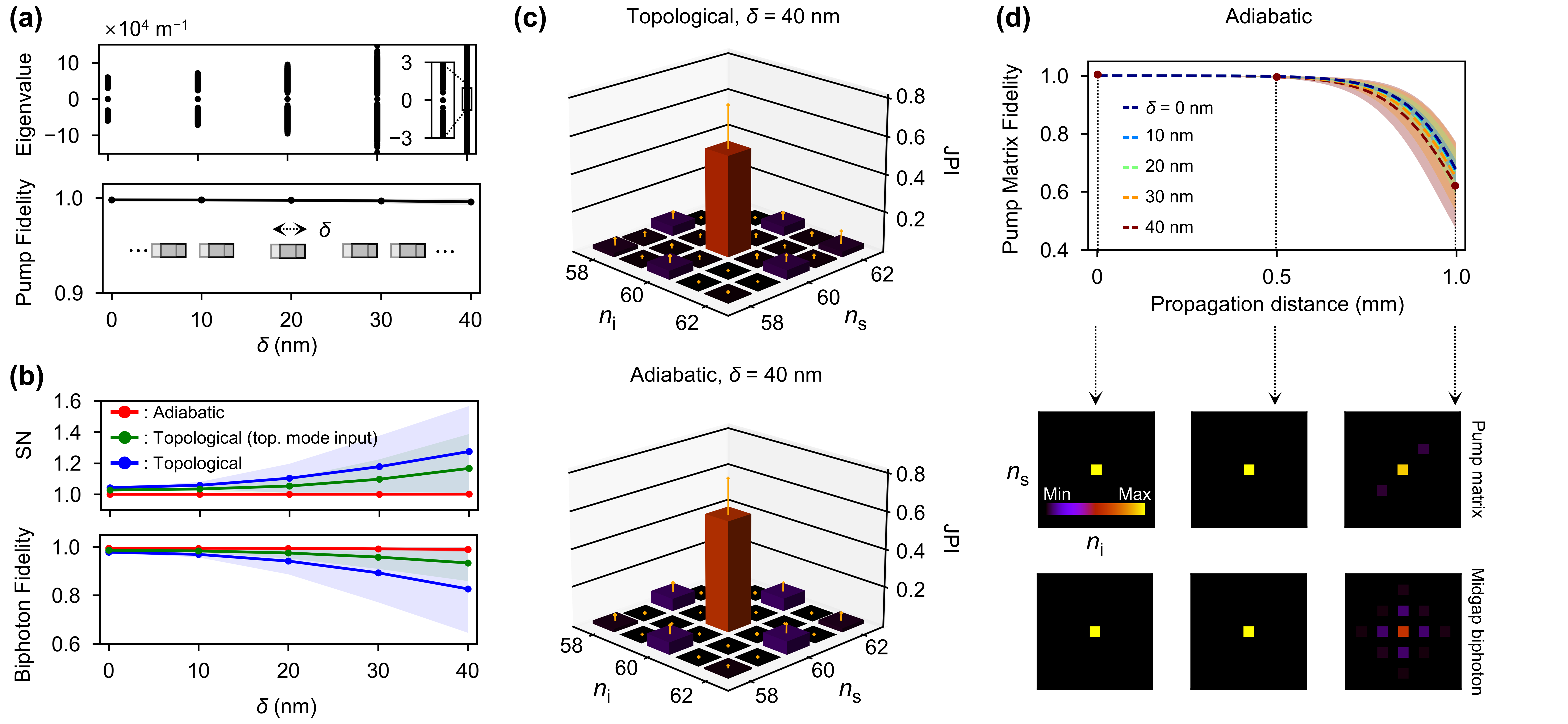}
  \caption{\label{fig3} Comparison of the gap disorder tolerance between an adiabatic structure and a conventional topological waveguide array.
    (a) Top: Eigenvalue spectra of the topological lattice samples across a range of gap disorder strength, \(\delta\) (\(0\) to \(40\,\mathrm{nm}\)).
    The inset confirms the persistence of the finite bandgap and midgap state even at maximum disorder.
    Bottom: Output pump fidelity of the adiabatic structure, maintaining near-unity values across all gap disorder levels.
    (b) Comparison of the output biphoton SN (top) and fidelity (bottom).
    The adiabatic scheme (red) exhibits superior resilience compared to the conventional topological array excited through a central waveguide input (blue) or the topological mode (green).
    (c) Ensemble-averaged output biphoton JPI for \(\delta = 40\,\mathrm{nm}\).
    The probability weight in odd-indexed waveguides for the topological array (top) indicates the admixture of non-topological biphoton modes, whereas the adiabatic structure (bottom) preserves the topological biphoton mode profile.
    The standard deviations are indicated by yellow error bars.
    (d) Fidelity of the normalized pump matrix relative to the midgap biphoton mode along the adiabatic passage for different disorder strengths.
    The nearly unchanged fidelity in the early stage illustrates the suppression of unintended nonlinear coupling within the strong isolation regime.
  The lower panels show the averaged squared-magnitude distributions of the pump matrix and the midgap biphoton mode at the highest \(\delta\).}
\end{figure*}

The spatial evolutions of the pump field and the generated biphoton state are obtained by numerically solving Eqs.~\eqref{eq2} and~\eqref{eq3}.
For the total device length of \(L=1.0\,\mathrm{mm}\), the propagation of the pump field is shown in Fig.~\ref{fig2}(a).
The pump field is initially matched to the midgap eigenmode of the isolated array, confining its intensity entirely within the central waveguide.
As it undergoes adiabatic evolution, the pump field acquires the spatial features of the topological mode, gradually spreading across the surrounding lattice.
To quantitatively characterize this mode evolution, we evaluate the fidelity, defined as $\left| \braket{T|\Psi} \right|^2 $, which measures the modal overlap between the propagating state $\ket{\Psi}$ and the target mode $\ket{T}$.
Fig.~\ref{fig2}(c) presents the fidelity between the evolving pump field and the midgap mode at each position throughout the device length.
The insets display the pump intensity profiles at the input and the output facets.
The fidelity is 1.0 at the input and 0.998 at the output, confirming a nearly perfect adiabatic transition into the topological mode.

Fig.~\ref{fig2}(b) shows the propagation of the generated biphoton probability distribution, where the two dimensional JPI is projected onto a single spatial dimension.
Unlike the pump, the biphoton is absent at the input and grows continuously via SFWM as the pump propagates along the device.
The biphoton state, initially confined to the isolated central waveguide, adiabatically transitions into the topological biphoton mode, distributing exclusively across the even-indexed waveguides.
As shown by the blue solid line in Fig.~\ref{fig2}(d), the fidelity between the propagating biphoton state and the midgap biphoton mode remains near-unity, varying only slightly from $1.0$ to $0.994$.
The insets depict the corresponding JPI and SN.
Here, the SN remains constant at $1.0$ along the entire propagation distance, indicating that the signal and idler photons are separable while their individual spatial profiles evolve toward the topological mode, mirroring the adiabatic behavior of the pump field.
The biphoton fidelity decreases slightly more than the pump fidelity because, for a separable state, it factorizes into the product of the signal and idler fidelities.
Since both individual fidelities are marginally below unity, their product inherently compounds the residual error.
Nevertheless, the output fidelity remains exceptionally high, confirming that the adiabatic transfer and topological biphoton generation proceed successfully with negligible degradation.

To investigate in detail how the pump generates the topological biphoton mode during the adiabatic passage, we analyze the nonlinear coupling within the eigenmode basis.
The biphoton state is expressed as $\ket{\psi} = \sum_{m,n} \bar{\psi}_{mn}\,\ket{m n}_{\mathrm{s,i}}$.
Here, $\ket{m n}_{\mathrm{s,i}}$ denotes the $m$-th and $n$-th eigenmodes of signal and idler, respectively, and $\left|\bar{\psi}_{mn}\right|^{2}$ is the biphoton population in this modal basis.
Eq.~\eqref{eq3} can be written as
\begin{equation}
  i\,\frac{\partial \bar{\psi}_{mn}}{\partial z}
  =
  (\beta_m^{\mathrm{s}} + \beta_n^{\mathrm{i}} )\bar{\psi}_{mn}
  + \gamma \sum_{j,k} A_j^{2}(z)\,\delta_{jk}(C_{mn})_{jk},
  \label{eq4}
\end{equation}
where $\beta_m^{\mathrm{s}}$ and $\beta_n^{\mathrm{i}}$ are the propagation constants of the signal and idler eigenmodes.
The coefficient \((C_{mn})_{jk}\equiv
(\ket{m}_{\mathrm{s}})_{j}^{*}(\ket{n}_{\mathrm{i}})_{k}^{*}\)
is the product of the complex conjugated signal and idler eigenmode amplitudes at waveguides \(j\) and \(k\), respectively.
The superscript asterisk denotes complex conjugation.
The resulting modal basis biphoton population is determined by
\begin{equation}
  \left|\bar{\psi}_{mn}(L)\right|^{2} =
  \gamma^{2}
  \left|
  \int_{0}^{L}
  \sum_{j,k} A_j^{2}(z)\,\delta_{jk}(C_{mn})_{jk}\,
  e^{i (\beta_m^{\mathrm{s}} + \beta_n^{\mathrm{i}}) z}\,dz
  \right|^{2}.
  \label{eq5}
\end{equation}

The term \(\sum_{j,k} A_j^{2}(z)\,\delta_{jk}(C_{mn})_{jk}\) in Eq.~\eqref{eq5} represents the SFWM modal overlap, where a larger magnitude indicates more efficient generation of the corresponding biphoton mode.
To quantify the selective nonlinear coupling to the target biphoton mode \(\ket{mn}_{\mathrm{s,i}}\), we define the normalized pump matrix fidelity \(\mathcal{F}_{mn}(z)\) as
\begin{equation}
  \mathcal{F}_{mn}(z) = \left|\sum_{j,k} \widetilde{A}_{j}^{2}(z) \delta_{jk}\,(C_{mn})_{jk} \right|^{2}.
  \label{eq6}
\end{equation}
Here, \(\widetilde{A}_{j}^{2}(z)=A_j^{2}(z)/[\sum_{\ell}|A_\ell^{2}(z)|^{2}]^{1/2}\), and \(\widetilde{A}_{j}^{2}(z)\delta_{jk}\) corresponds to the normalized pump matrix.
The blue dashed line in Fig.~\ref{fig2}(d) tracks this fidelity relative to the midgap biphoton mode; at the initial stage, the isolation gap ensures tight spatial confinement of both the pump and the midgap biphoton mode within the central site.
This leads to unit fidelity and facilitates selective generation of the midgap biphoton mode.
Consequently, the biphoton state generated in the initially isolated region is successfully guided through the adiabatic link to reach the topological biphoton mode.

We next evaluate the robustness of the proposed adiabatic scheme.
The fabrication imperfection considered here is gap disorder, i.e., random variations in the separations between adjacent waveguides.
This gap disorder is implemented by adding independent Gaussian noise with standard deviation \(\delta\), ranging from \(0\) to \(40\,\mathrm{nm}\) in \(10\,\mathrm{nm}\) increments.
Because the gap disorder preserves the chiral symmetry and the bulk bandgap remains open for the disorder strengths considered, the topological invariant is unchanged.
We generate 300 random samples for each disorder strength and simulate them numerically.
Furthermore, for direct comparison, we simulate a conventional topological waveguide lattice under identical disorder conditions.
This conventional topological waveguide lattice is constructed by fixing its geometry to the output cross section of the adiabatic structure over the entire device length.

Fig.~\ref{fig3}(a) shows the eigenvalue spectra of the topological array at various gap disorder levels, $\delta$.
While the bandgap narrows as \(\delta\) increases, the inset confirms that a finite gap persists even at the maximum disorder strength, verifying the presence of the protected mode.
For the adiabatic structure, the lower panel of Fig.~\ref{fig3}(a) shows the fidelity between the output field and the topological mode of each disordered sample when the pump is injected into the central waveguide.
Each point represents the mean fidelity, and the shaded region indicates one standard deviation.
The mean fidelity remains near unity, reaching 0.996 at the highest disorder level and confirming the robustness of the adiabatic pump evolution.

Biphoton generation is simulated with the pump injected into the central waveguide for both the adiabatic and conventional topological structures.
To avoid pump mode beating in the topological array and to assess the effect of disorder on the nonlinear interaction, we also consider launching the pump into the topological mode.
The fidelity of the output biphoton is calculated with respect to the topological biphoton mode for each disordered sample, serving as a quantitative measure of how faithfully the target mode is generated.
Additional analyses that consider more practical conditions---such as the fidelity relative to the ideal topological biphoton state and the influence of waveguide width imperfections---are provided in the Supporting Information.

Fig.~\ref{fig3}(b) presents the SN and fidelity of the output biphoton state.
In the conventional topological array with central waveguide injection, the SN increases to \(1.27\pm0.293\) while the fidelity drops to \(0.826\pm0.181\) at the highest disorder strength.
Even when the pump is launched into the topological mode, the SN still rises to \(1.17\pm0.221\) and the fidelity degrades to \(0.933\pm0.074\), accompanied by a relatively broad distribution.
This confirms that biphoton generation in the conventional topological lattices is not fully resilient to disorder, consistent with a prior study~\cite{bergamasco2019generation}.
In contrast, the adiabatic structure preserves the near-unity SN, while the fidelity exhibits only a slight reduction to 0.989 with a narrow distribution, demonstrating its superior tolerance to disorder.
The increase in SN observed in the topological array signifies the admixture of non-topological components into the output biphoton state.
This is further evidenced by Fig.~\ref{fig3}(c), which presents the ensemble-averaged JPI together with the standard deviation for the highest \(\delta\).
Unlike the adiabatic structure, the topological array shows probability weight in odd-indexed waveguides, clearly indicating contributions from non-topological biphoton modes.

\begin{figure*}[t]
  \centering
  \includegraphics[width=1\textwidth]{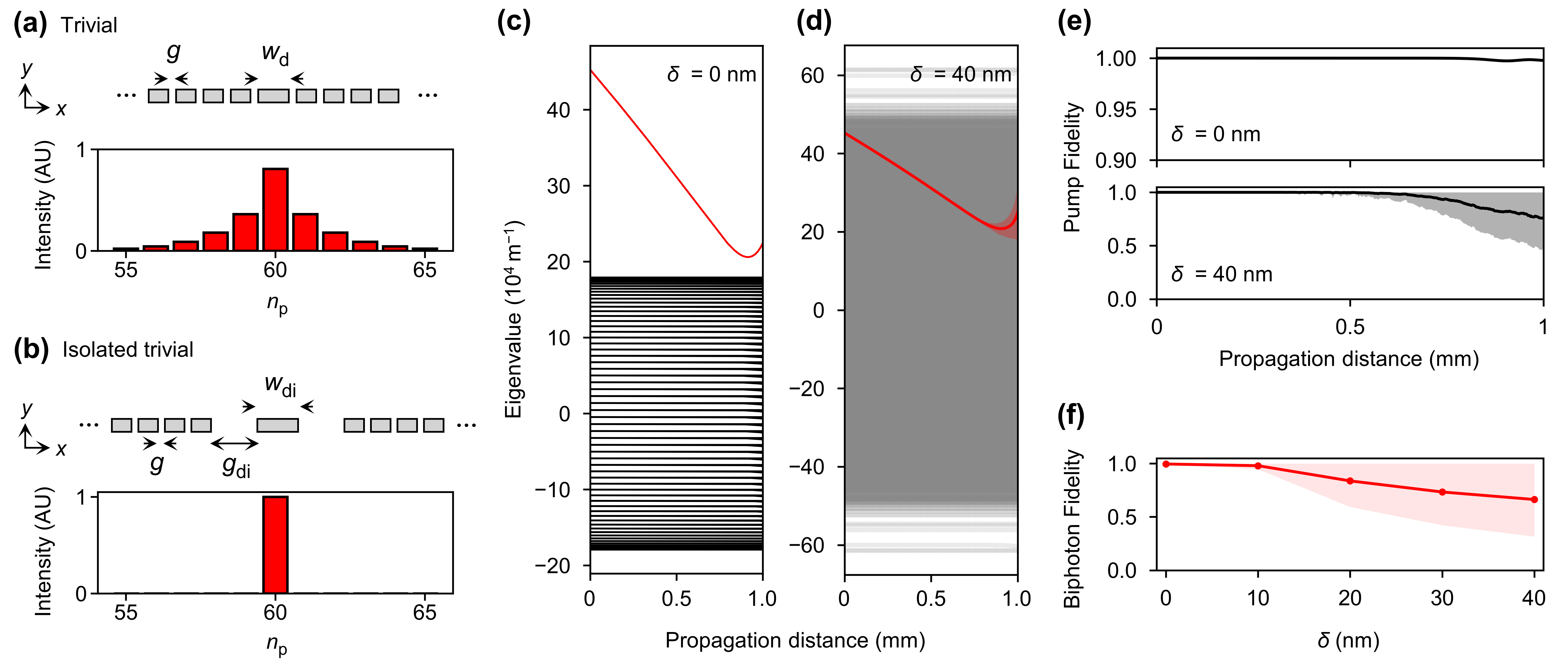}
  \caption{\label{fig4}
    Adiabatic scheme for a trivial defect case.
    (a) Cross section of the trivial defect array (top) and the intensity profile of its defect mode (bottom).
    (b) Cross section of the isolated trivial defect array (top) and its corresponding defect mode intensity profile (bottom).
    (c) Eigenvalue spectrum along the propagation direction in the absence of disorder.
    The defect mode (red curve) maintains spectral separation from the bulk bands (black lines) throughout the transition.
    (d) Eigenvalue spectrum at $\delta = 40\,\mathrm{nm}$.
    The gray regions show the bulk eigenvalues, while the red curve and shaded band denote the mean and standard deviation of the defect mode eigenvalue.
    (e) Pump field fidelity with respect to the defect mode along the propagation without disorder (top) and for $\delta = 40\,\mathrm{nm}$ (bottom).
  (f) Output biphoton fidelity as a function of the gap disorder strength $\delta$.}
\end{figure*}

\begin{figure}[t]
  \centering
  \includegraphics[width=0.5\textwidth]{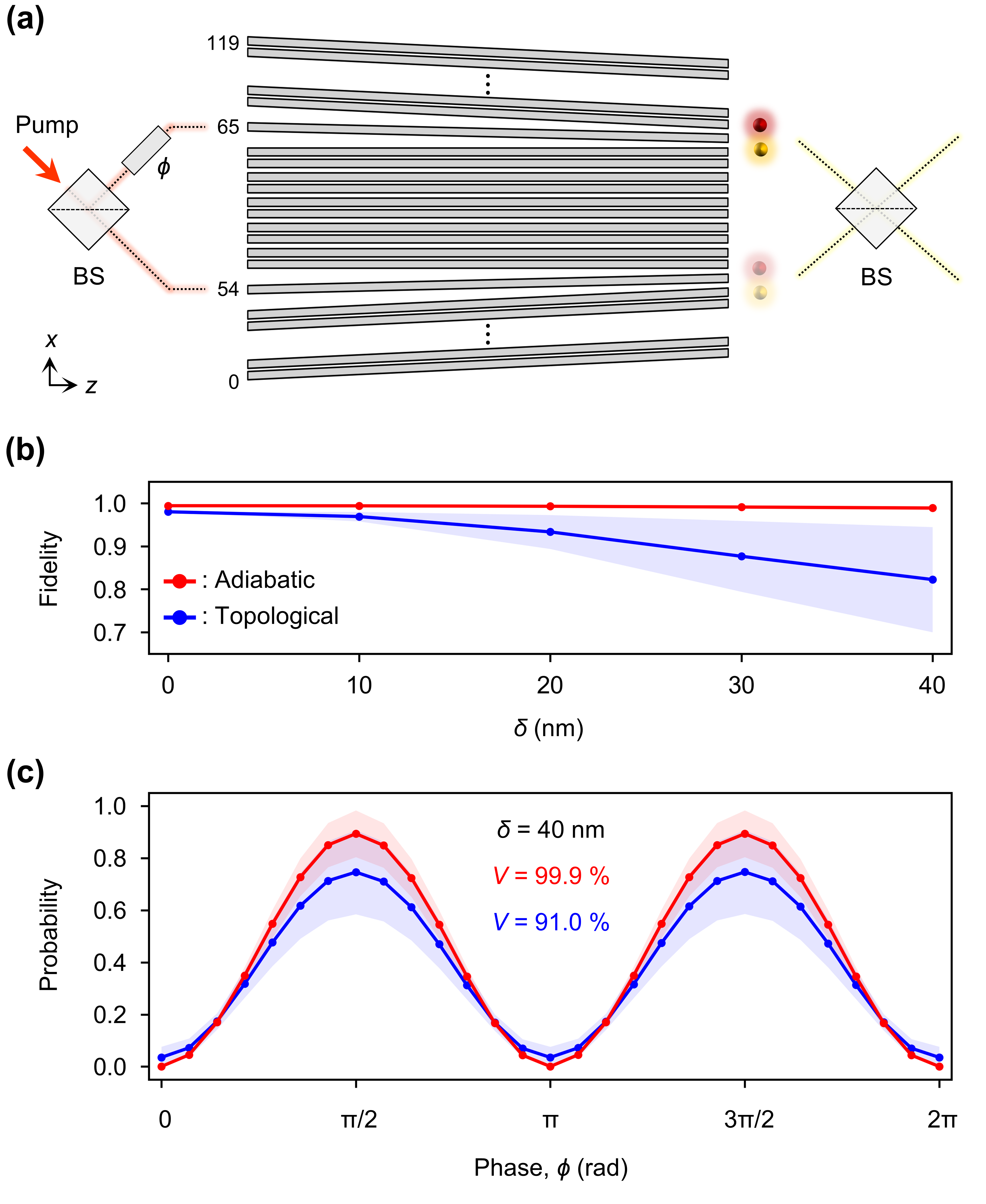}
  \caption{\label{fig5} Robustness of path entangled topological biphoton mode generation and quantum interference in the presence of gap disorder.
    (a) Schematic of an adiabatic passage for the generation of path entangled topological biphoton states. An isolated-waveguide array containing two isolated sites is adiabatically connected to an SSH lattice hosting two corresponding topological defects at sites 54 and 65.
    (b) Fidelity of the generated biphoton state relative to the NOON state for $\phi = 0$. The biphoton states generated in the adiabatic structure (red) are compared with those generated in the topological lattice (blue).
  (c) Probability of the split mode after quantum interference at an output BS as a function of the pump phase delay, $\phi$. The gap disorder strength is \(40\,\mathrm{nm}\), and the visibilities are indicated.}
\end{figure}

The enhanced robustness of our structure arises from the complementary roles of strong isolation and the disorder-tolerant adiabatic evolution enabled by the topological mode.
Fig.~\ref{fig3}(d) shows the mean fidelity of the normalized pump matrix relative to the midgap biphoton mode along the adiabatic passage for different disorder strengths.
In the early stages of propagation, the isolation gap ensures that the pump matrix fidelity remains unchanged against disorder, effectively suppressing unintended nonlinear coupling to the non-topological biphoton mode.
The lower panels of Fig.~\ref{fig3}(d) present the averaged squared-magnitude distributions of both the pump matrix and the midgap biphoton mode at the input, intermediate, and output positions for \(\delta = 40\,\mathrm{nm}\).
In the strong isolation region, each distribution remains strictly confined to the central waveguide.
As the pump transitions into the topological mode, disorder-induced spatial deformation reduces the mean fidelity and increases the variance.
This regime is analogous to injecting the topological mode directly into the topological array, where the spatially extended profile is deformed by disorder, degrading the modal overlap.
In our strategy, however, this degradation is mitigated during the strong isolation stage and adiabatic evolution, thereby enhancing the overall robustness.

To further clarify the origin of the enhanced robustness, we apply the adiabatic scheme to a waveguide array supporting a trivial defect mode and compare its disorder tolerance with that of the topological case.
As shown in Fig.~\ref{fig4}(a), the trivial defect array has uniform waveguide gaps of $g = 100\,\mathrm{nm}$, with the central waveguide width increased to $w_\mathrm{d} = 465\,\mathrm{nm}$ to form a localized defect mode.
These parameters are chosen so that $97.8\%$ of the mode intensity is confined within the central five waveguides, closely matching the $98.1\%$ confinement of the topological defect mode shown in Fig.~\ref{fig1}(d).
The corresponding isolated array (Fig.~\ref{fig4}(b)) has a central width of $w_\mathrm{di} = 500\,\mathrm{nm}$ and isolation gaps of $g_\mathrm{di} = 800\,\mathrm{nm}$.
The two arrays are connected by linearly varying both the gaps and the central waveguide width over a device length of $1\,\mathrm{mm}$.

Fig.~\ref{fig4}(c) shows the evolution of the eigenvalue spectrum along the propagation axis in the absence of disorder.
By appropriately tapering the central waveguide during the transition, the defect mode eigenvalue (red curve) avoids crossing into the bulk bands, maintaining a finite spectral separation throughout the entire connection.
Consequently, a pump injected into the central waveguide undergoes near-ideal adiabatic evolution, reaching an output fidelity of $0.998$, which is comparable to that of the topological case.

The critical difference between the trivial and topological cases emerges when gap disorder is introduced.
Fig.~\ref{fig4}(d) shows the eigenvalue spectrum of the trivial defect structure at the disorder strength of $\delta = 40\,\mathrm{nm}$, where the gray regions are the bulk eigenvalues of all $300$ samples and the red curve with the shaded region denotes the mean and standard deviation of the defect mode eigenvalues.
In contrast to the topological case (Fig.~\ref{fig3}(a)), the spectral separation between the trivial defect mode and the bulk bands is not preserved.
This breakdown directly degrades the adiabatic transfer.
As shown in Fig.~\ref{fig4}(e), while the structure exhibits nearly ideal adiabatic transfer, gap disorder disrupts the evolution and significantly lowers the fidelity.
This degradation is also reflected in the output biphoton fidelity (Fig.~\ref{fig4}(f)), which falls rapidly as disorder increases.
This control case confirms that strong spatial localization alone cannot guarantee disorder-tolerant biphoton generation.
The topological mode is essential in the proposed design because its spectral isolation is required for the adiabatic transfer under gap disorder.

Beyond single-mode generation, the nonlinear SSH lattice can host entangled topological biphoton states by increasing the number of defect waveguides~\cite{blanco2018topological, wang2019topologically}.
By splitting a weak pump using a beam splitter (BS) and coherently injecting it into two defect sites, a path entangled NOON state can be prepared.
The generation of path entangled NOON states and their multiphoton interference in on-chip devices provide useful building blocks for quantum information processing~\cite{silverstone2014chip, wang2016chip, maeder2026programmable}.
We investigate whether the adiabatic passage makes this entangled state generation robust against gap disorder.
Fig.~\ref{fig5}(a) shows an adiabatic passage from an array with two isolated sites to a topological lattice hosting two corresponding defects.
Each defect supports the topological mode, and the two modes are spatially separated by ten waveguides to suppress coupling.
The structure parameters and the gap disorder are the same as before.
The results are compared with a conventional topological array identical to the output of the adiabatic passage under the same pump injection conditions.

We denote the topological modes supported at sites 54 and 65 as $\ket{\alpha}$ and $\ket{\beta}$, respectively.
A phase delay $\phi$ is applied to one arm, and the resulting SFWM process generates the NOON state:
\begin{equation}
  \frac{1}{\sqrt{2}}
  \left(
    \ket{\alpha}_s \ket{\alpha}_i
    +
    e^{2 i \phi}
    \ket{\beta}_s \ket{\beta}_i
  \right),
  \label{eq7}
\end{equation}
where the $2\phi$ phase originates from the simultaneous annihilation of two pump photons.
To evaluate the quality of the generated entangled state, we simulate quantum interference through an output BS~\cite{blanco2018topological, wang2019topologically}.
The BS operation is described by $a^{\dagger\,\text{in}}_j=\frac{1}{\sqrt{2}}\left(a^{\dagger\,\text{out}}_j\pm a^{\dagger\,\text{out}}_{N-1-j}\right)$, where the minus (plus) sign applies for $j < N/2$ ($j \ge N/2$).
Here, $N$ is the total number of waveguides, and $a^\dagger_j$ denotes the creation operator at site $j$.
The BS transmits the $j$-th waveguide mode to the $(N-1-j)$-th mode and reflects it back to the $j$-th mode.
As a result, the BS operation transforms the state into a superposition of bunched and split modes:
\begin{equation}
  \cos\phi \, \ket{\psi_{\text{bunch}}}
  +
  i \sin\phi \, \ket{\psi_{\text{split}}},
  \label{eq8}
\end{equation}
where
\begin{align}
  \ket{\psi_{\text{bunch}}}
  &=
  \frac{1}{\sqrt{2}}
  \left(
    \ket{\alpha}_s \ket{\alpha}_i
    +
    \ket{\beta}_s \ket{\beta}_i
  \right), \\
  \ket{\psi_{\text{split}}}
  &=
  \frac{1}{\sqrt{2}}
  \left(
    \ket{\alpha}_s \ket{\beta}_i
    +
    \ket{\beta}_s \ket{\alpha}_i
  \right).
  \label{eq9}
\end{align}
The bunched mode describes a state in which the signal and idler occupy the same defect mode, whereas the split mode corresponds to spatially separated pairs.

To assess whether the adiabatic structure robustly generates the path entangled topological biphoton mode, we compute the output biphoton fidelity when $\phi = 0$.
Fig.~\ref{fig5}(b) shows the fidelity as a function of disorder strength.
The adiabatic structure maintains the fidelity close to unity, whereas the conventional topological array exhibits a strong degradation.
As in the single defect case, the conventional array generates unintended spatial mode components.

The robustness of the adiabatic passage is further evidenced by the interference fringes of the split mode probability in Fig.~\ref{fig5}(c).
The disorder strength is fixed at \(40\,\mathrm{nm}\), and the pump phase delay is varied from $0$ to $2\pi$.
To quantify the visibility $V$, we compute
\begin{equation}
  V
  =
  \frac{P_{\max}-P_{\min}}{P_{\max}+P_{\min}},
\end{equation}
where $P_{\max}$ and $P_{\min}$ are the maximum and minimum mean split mode probabilities, respectively.
Both structures exhibit a $\pi$-periodic oscillation consistent with the expected $\sin^2\phi$ dependence.
The adiabatic structure achieves an exceptional visibility of $99.9 \%$ and a maximum split mode probability of $89.3 \%$.
The reduction in the maximum probability arises from disorder-induced deformations in the $\ket{\alpha}$ and $\ket{\beta}$ mode profiles, which hinder complete cancellation of the bunched mode component.
In contrast, the conventional topological waveguide array shows a lower visibility of $91 \%$ and a peak probability of only $74.7 \%$, limited by the admixture of unintended spatial components.
These results demonstrate that adiabatic passage enables robust quantum interference using biphoton entanglement even in the presence of significant waveguide gap imperfections.

\section{\label{sec3}Conclusion}
In conclusion, we have proposed and numerically demonstrated robust generation of topological biphoton modes in the nonlinear SSH waveguide lattice via adiabatic passage.
While conventional topological waveguide arrays suffer from excitation of non-topological biphoton states under structural disorder, the adiabatic passage effectively mitigates this degradation.
By initiating the nonlinear process in the strong isolation regime, generation of unintended biphoton modes is suppressed.
The generated biphoton state is subsequently transferred into the topological biphoton mode with high fidelity through adiabatic evolution.
Numerical simulations show that the adiabatic structure maintains near-unity output biphoton fidelity and SN under substantial gap disorder, in clear contrast to the degradation observed in conventional topological waveguide arrays.
Through the comparative study with the trivial defect array, we confirm that this exceptional robustness arises from the synergistic combination of strong spatial isolation and the topological mode within the adiabatic passage.
This strategy further extends to path entangled NOON states, which exhibit near-unity quantum interference visibilities despite realistic waveguide gap disorder.
Our work provides a practical route toward disorder-tolerant, high fidelity quantum state preparation in integrated photonics platforms.

\begin{acknowledgments}
  This work was supported by the Agency for Defense Development Grant funded by the Korean Government.
\end{acknowledgments}

\end{document}